\documentstyle[12pt]{article}
\author{Hans - J\"urgen Schmidt}
\title{Conformal relations and Hamiltonian formulation of 
fourth--order gravity\footnote{Extended 
version of a lecture 
read at the   International School-Seminar 
``Problems of Theoretical
Cosmology'', 1-7 September 1997, Ulyanovsk, Russia}}
\date{}
\begin{document}
\maketitle

\renewcommand{\baselinestretch}{1.2}

\medskip

\centerline{University Potsdam, Institute for Mathematics, 
 D-14415 POTSDAM} 

\centerline{PF 601553, Am Neuen Palais 10, Germany
  \  \   hjschmi@rz.uni-potsdam.de} 
 
\medskip

 Grav. and Cosm., to appear Febr. 1998

\medskip

\begin{abstract} 
 The conformal equivalence of
fourth--order gravity following from a
non--linear Lagrangian $L(R)$
 to theories of other types is widely known, here 
we report on a new 
conformal equivalence of these theories to theories of the
same
type but with different Lagrangian.

For a quantization of fourth-order theories
 one needs a Hamiltonian formulation of them.
 One of the possibilities to do so goes back to
 Ostrogradski in 1850. Here we present another 
possibility: 
 A  Hamiltonian $H$ different from Ostrogradski's one is
discussed
for the Lagrangian $L=L(q, \dot q, \ddot q)$,
where $\partial \sp 2 L / \partial (\ddot q)\sp 2 \ne 0 $. We add
a suitable divergence to $L$ and insert $a=q$ and
 $b=\ddot q$. Contrary to other approaches no constraint is
needed because $\ddot a = b$ is one of the canonical equations.
Another canonical equation becomes equivalent to the
fourth--order Euler--Lagrange equation of $L$.

Finally, we discuss the stability properties 
 of cosmological models within fourth--order gravity.
\end{abstract}

\medskip

discrete systems

\bigskip

\section{Introduction}

The purpose of this review is to present some
of the points  which have been discussed in
 recent years in connection with cosmological 
models following from fourth--order theories of gravity.

Usually, one applies some
 version of a  conformal equivalence theorem
between these theories and Einstein's theory with 
additional scalar fields. But there exists 
also another type of conformal equivalence: 
 In  [1], we have shown
 that for non--linear $L=L(R)$, $G=dL/dR \ne 0$ the
Lagrangians
$L$ and $ \hat L ( \hat R)$ with
$\hat L  =  2R/G\sp 3  -  3 L/G \sp 4 $, 
 $\hat g_{ij} = G \sp 2 \, g_{ij}$ and
$\hat R =  3R/G\sp 2  -  4 L/G \sp 3 $ 
give conformally equivalent fourth--order field equations
being
dual to each  other. The proof represents a new
application of the fact that the
operator $\Box - \frac{R}{6}$ is conformally invariant.

The Ostrogradski approach [2] to find a Hamiltonian formulation
for a higher--order theory is the most famous 
(see  [2-9]) but  not the only method. In sct. 2
we present an alternative Hamiltonian formalism for fourth--order
theories. It systematizes what has been sporadically
done in the literature for special examples.

     Sct. 3 deals with fourth--order gravity following from a
non--linear Lagrangian $L(R)$.  The instability of
these theories from the point of view of the Cauchy problem is
discussed.

Sct. 2 applies to general higher--order 
 theories, sct. 3 to gravity, and both
are combined to fourth--order cosmology in sct. 4. We
discuss known exact solutions  under the
stability criteria mentioned before.

In the final sct. 5 we  list some open problems
for future research.

     \ The rest of this intro\-duction short\-ly re\-views
pa\-pers on higher--or\-der theo\-ries.  Eliezer and Woodard [2]
and  Jaen, Llosa, Molina and Vives [3]
represent  standard papers for the generalization of the
Ostrogradski approach to non--local systems  and to systems with
constraints (see also [4-6]) applying Dirac's approach. 

      Let the Lagrangian $L$ be  a function of the vector
$q_{\alpha}$ and its first $n$ temporal derivatives
$\dot q_{\alpha}$, $\ddot q_{\alpha}$, 
$\dots , \,  q^{(n)}_{\alpha}$.
The Hessian is
\begin{equation}
H_{\alpha \beta}=\frac{\partial^2L}{\partial q_{\alpha}
^{(n)}\partial q_{\beta}^{(n)}}
\end{equation}
and the non--vanishing of its determinant 
defines the regularity of $L$. In the following we do not write
the subscript $\alpha$; one can think of $q$ as being a  point
particle in a (one-- or higher--dimensional) space. In the
Ostrogradski approach, $Q = \dot q$ is taken as additional
position variable. This leads to an ambivalence of the procedure,
because it is not trivial to see at which places $\dot q$ has to
be replaced with $Q$, cf. [7]. We prevent this ambivalence in our
alternative Hamiltonian, cf. sct. 2, by putting $Q=\ddot q$.

Ref. [8] discusses higher--order field theories. The problem is
the lack of an energy bound, typically two kinds of oscillators
with different signs of energy exist. Usually, one restricts the
space of initial conditions to prevent negative energy solutions.
The authors of ref. [8]  redefine the energy analogous to the
Timoshenko model, so one gets a positive mechanical energy
inspite of an indefinite Ostrogradski Hamiltonian, they write:
``An appealing aspect of this approach is the absence of any
constraint.'' So it has this property in common with our approach
sct. 2, but it is otherwise a different one. Another standard
procedure [3,8,9] for deal\-ing with 
higher--or\-der Lagran\-gians is to con\-si\-der them as a
se\-quence in a para\-meter $\epsilon$, so one can break the
Euler--Lagrange--equa\-tion into a se\-quence of se\-cond or\-der
ones. In [9] this is called ``reduction of higher--order
Lagrangians by a formal power series in an ordering parameter.''
[9] deals also with the Lie--K\"onigs theorem: a local
Hamiltonian is always possible,  and they consider  some global
questions.

     \ \ Let us show the famous counter--example [10]: 
it is an example of a second order system not following
from a Lagrangian. 
$$\ddot x + \dot y =0 \quad \quad \ddot y + y + \epsilon \dot x
=0$$
It follows from the Lagrangian
$$L=\frac{1}{2} [\dot y^2 - y^2 + \epsilon(
x \dot y - \dot x y - \dot x^2 )]$$
for $\epsilon \ne 0$ and has no Lagrangian otherwise. We mention
this example to show that the following recipe need 
not to work always. Recipe for higher--order theories: ``Write
down the Euler--Lagrange equations, break them into a sequence of
second order ones by introducing further coordinates. Find
Lagrangians for these second order equations.''

A powerful method for dealing with  a classical Lagrangian 
\begin{equation}
L=\frac{1}{2} \, g_{ij} \, \dot q^i \, \dot q^j - V(q)
\end{equation}
is given in [11]. The Euler--Lagrange
equation  to Lagrangian (1.2) reads 
$$
\ddot q^i + \Gamma ^i _{jk} \dot q^j \dot q^k 
= - g^{ik} V _{,k} 
$$ 
and is fulfilled for geodesics in the Jacobi--metric
$$ \hat g_{ij} \ = \ (E-V) \, g_{ij}$$ 
Remark: For constant potentials $V$ this is trivial, for
non-constant potentials the constant $E$ must be correctly chosen
to get the result, for $E=V$ it breaks down, of course.

  \ \  Stelle [12] cites Ostrogradski [2] but uses other methods
to extract different spin modes for fourth--order  gravity. In
[13], a regular reduction of fourth--order gravity similar to the
method with an ordering parameter mentioned above has been
proposed as follows:  In the Newtonian limit one has 
$$\Delta \Phi + \beta \Delta
\Delta \Phi = 4\pi G \rho ,$$
then one restricts to solutions which can be expanded into powers
of the coupling parameter $\beta $. Argument: If $\beta $ is a
parameter, this is well justified, if it is a universal constant,
then this restriction is less satisfying.
Comment: This restriction excludes the usual Yukawa--like
potential $\frac{1}{r} \exp (-r/\sqrt \beta )$, so one may  doubt
whether this method gives the right solutions. Let us further
mention ref. [14] for non--local gravitational Lagrangians like
$L=R\Box^{-1}R + \Lambda$ in two dimensions and refs. [15-18]
for the linearized $R^2$--theory.

To facilitate the reading of sct. 2, we pick up the example  
 eq. (5) of [5]:
\begin{equation}
\tilde L = [ \ddot q \sp 2 + 4 \ddot q \dot q \sp 2 + 4 \dot q
\sp 4 ] e \sp{3q}
\end{equation}
The equation of motion is [5, eq. (6)]
\begin{equation}
2 q \sp{(4)} + 12 \dot q q \sp{(3)} + 9 ( \ddot q )\sp 2 + 18
\dot q \sp 2 \ddot q = 0
\end{equation}
A good check of the validity of the formalism is the following:
For a constant $c>0$ and $\dot q>0$, each solution of 
$$\ddot q = - 2 \dot q \sp 2 + c \sqrt{ \dot q }$$
is also a solution of eq. (1.4). 

 By adding a divergence to eq. (1.3) one gets $L = (\ddot q)\sp 2
e\sp {3q}$. The alternative formalism requires to use $q\sp 1 = q
$ and $q \sp 2 = \ddot q $ as new coordinates. So we get
\begin{equation}
L = (q\sp 2)\sp 2 \ \exp (3 q\sp 1)
\end{equation}
Eq. (1.5) represents the ultralocal Lagrangian mentioned in [5].
It is correctly stated in [5], that the alternative  formalism 
does not work for this version  eq. (1.5) of the system. This 
clarifies that the addition of a divergence to a higher--order
Lagrangian  sometimes influences the 
applicability of the  alternative Hamiltonian formalism. 
So one should add a ``suitable''  total derivative to the
Lagangian. ``Suitable''  means, that the space of 
solutions is the same at both sides, and that the relation
between the various coordinates is ensured without imposing any
constraints. It turns out, that the Lagrangian $\hat L$ 
differing from $\tilde L$, eq. (1.3) by a divergence only
\begin{equation}
\hat  L = - [ \ddot q \sp 2 + 6 \ddot q \dot q \sp 2 + 2 \dot q
q\sp {(3)} ] e \sp{3q}
\end{equation}
does the job. Of course, the variations of $L$, $\tilde L$, and
$\hat L$ with respect to $q$ all give the same equation of motion
(1.4). But only in version (1.6) the alternative  formalism
(insertion of the equation $Q=\ddot q$ and then apply the usual
formulas of classical mechanics - to avoid ambiguities with the
square--sign we have replaced $q^1$ by $q$ and $q^2$ by $Q$)
leads correctly to the Hamiltonian [5, eq. (7)]:
\begin{equation}
H=-\frac{1}{2}(pP - 3 P^2 Q)e^{-3q} + Q^2 e^{3q}
\end{equation}
It essentially differs from the Ostrogradski approach because
terms only linear in the momenta do not appear.
 The integrability condition $Q=\ddot q$ and the equation of
motion (1.4) both follow from the canonical equations of eq.
(1.7); no constraint is necessary to get this.

\medskip

Wagoner [19] showed conformal relations between
different types of second order theories 
(Einstein, Brans-Dicke), whereas refs. [20-23] discussed such
relations between fourth--order and second order theories.
  Refs. [19-23] also discuss which of the conformally equivalent
metrics can be considered physical. 

\medskip

\section{The alternative Hamiltonian formalism}
\setcounter{equation}{0}

Let us consider the Lagrangian
\begin{equation}
L \ = \ L(q, \dot q , \ddot q)
\end{equation}
for a point particle $q(t)$, a dot denoting $\frac{d}{dt}$ and
$$q\sp{(n)} =  \frac{d\sp n q}{dt\sp n}$$
The corresponding Euler--Lagrange  equation reads
\begin{equation}
0 = \frac{\partial L}{\partial q} - \frac{d}{dt}
\frac{\partial L}{\partial \dot q} +   \frac{d\sp 2}{dt \sp 2}
\frac{\partial L}{\partial \ddot q}
\end{equation}
We suppose this Lagrangian to be non-degenerated, i.e., $L$ is
non-linear in $\ddot q $. The highest-order term of eq. (2.2) is 
$$q\sp{(4)} \frac{\partial \sp 2 L}{\partial (\ddot q) \sp 2}$$
therefore, non-degeneracy (= regularity, cf. eq. (1.1)) is
equivalent to require that eq. (2.2) is of fourth order, i.e.
$$ \frac{\partial \sp 2 L}{\partial (\ddot q) \sp 2} \ne 0$$
(If $q$ is a vector consisting of $m$ real components then this
condition is to be written as Hessian determinant.)

 If we add the divergence 
$\frac{d}{dt} G(q, \dot q)$ to $L$, we do not alter the
Euler--Lagrange equation (2.2). Furthermore, the expression
$\frac{d}{dt} G$ is linear in $\ddot q$, and so its addition to
$L$ does not influence the condition of non--degeneracy. The
addition of such a divergence can therefore simply absorbed by a
suitable redefinition of $L$. 

     In the next two subsections we add a special and a more
general divergence to get a Hamiltonian formulation different
from Ostrogradski's one. In  Kasper [4] a similar
consideration has been made at the Lagrangian's  level.
Subsection 2.1 represents only a special case of subsection 2.2,
but we write it down, because it has the advantage that the
formulas can be given explicitly, and so the formalism becomes
more transparent.  

\subsection{A special divergence}
     The addition of the following divergence is no more done by
a redefinition of $L$ 
\begin{equation}
L_{div} = \frac{d}{dt}[ f(q) \, \dot q \, \ddot q], \ \ f(q) \ne
0
\end{equation}
and we consider $\hat L = L + L_{div}$. The Euler--Lagrange
equation is again eq. (2.2). Using 
$$f'(q) \equiv \frac{df}{dq}$$
we get 
\begin{equation}
\hat L = L + f'(q)\dot q \sp 2 \ddot q +
f(q)[(\ddot q)\sp 2 + \dot q q \sp{(3)} ]
\end{equation}
which contains third derivatives of $q$.

     We introduce new coordinates
\begin{equation}
a \ = \ q\, , \ b \ = \ \ddot q
\end{equation}
(In the Ostrogradski approach, the second coordinate is $\dot q$,
instead.)
It is obvious that there is exactly this one compatibility
condition:
\begin{equation}
\ddot a \ = \ b
\end{equation}
Let us insert eq. (2.5) into eq. (2.4). This insertion becomes
unique by the additional requirement that $\hat L $ does not
depend on second and higher derivatives of $a$ and $b$, i.e.,
$$\hat L \ = \  \hat L (a, \dot a, b, \dot b ) $$
giving
\begin{equation}
\hat L = L(a, \dot a, b) + f'(a) \dot a \sp 2 b +
f(a)[b\sp 2 + \dot a \dot b ]
\end{equation}
(In the Ostrogradski approach, there remains an ambivalence which
of the $\dot q$ in the original Lagrangian is to be interpreted
as second coordinate and which as time derivative of the first
one.)
     
     The momenta are defined as in classical mechanics by
\begin{equation}
p_a = \frac{\partial \hat L}{\partial \dot a}, \ 
p_b = \frac{\partial \hat L}{\partial \dot b}
\end{equation}
(In the Ostrogradski approach, an additional term is necessary.)
Inserting eq. (2.7) into eqs. (2.8) we get
\begin{equation}
p_a = \frac{\partial  L}{\partial \dot a}
+ 2 f'(a) \dot a b + f(a)\dot b
\end{equation}
and
\begin{equation}
p_b = f(a) \dot a
\end{equation}
Because of $f(a) \ne 0 $, cf. eq. (2.3), we can invert eq. (2.10)
to
\begin{equation}
\dot a = \frac{p_b}{f(a)}
\end{equation}
Inserting eq. (2.11) into eq. (2.9) and dividing by $f(a)$ we get
\begin{equation}
\dot b = \frac{1}{f(a)}[p_a - \frac{\partial  L}{\partial \dot a}
- 2 f'(a) b \frac{p_b}{f(a)}]
\end{equation}
It is instructive to make a more general consideration: The
question, whether eqs. (2.9, 10) can be inverted to $\dot a$,
$\dot b$, can be answered by calculating the Jacobian
 (by the way: Carl Gustav Jacob Jacobi was born in
Potsdam 10th December 1804.)
\begin{equation}
J = \frac{\partial (p_a, p_b)}{\partial (\dot a, \dot b)}
= \frac{\partial p_a}{\partial \dot a}
\frac{\partial p_b}{\partial \dot b}
- \frac{\partial p_a}{\partial \dot b}
\frac{\partial p_b}{\partial \dot a}
\end{equation}
We insert eqs. (2.9, 10) into eq. (2.13) and get 
\begin{equation}
J \ = \ - \ [f(a)]\sp 2
\end{equation}
Because of $f \ne 0$ one has also $J \ne 0$ and the inversion is
possible. This more general consideration gave the additional
information that the Jacobian is always negative; this is one of
the several possibilities to
give the phrase ``fourth--order theories 
are  unstable'' a concrete meaning.

     We define the Hamiltonian $H$ as usual by
$$H \ = \ \dot a p_a + \dot b p_b - \hat L$$
i.e., with eq. (2.7) we get
\begin{equation}
H  =  \dot a p_a + \dot b p_b -  L - f'(a) \dot a \sp 2 b -
 f(a) [ b\sp 2 + \dot a \dot b ]
\end{equation}

Here we insert $\dot a$ according to eq. (2.11) and get the
Hamiltonian $H \ = \ H(a, p_a, b, p_b)$. 
The factor of  $\, \dot b$ in $H$ automatically vanishes, so we
do not need eq. (2.12).  
The canonical equations read
\begin{equation}
\frac{\partial H}{\partial p_a} \ = \ \dot a
\end{equation}
further
\begin{equation}
\frac{\partial H}{\partial p_b} \ = \ \dot b
\end{equation}
and
\begin{equation}
\frac{\partial H}{\partial a} \ = \ - \dot p_a
\end{equation}
and
\begin{equation}
\frac{\partial H}{\partial b} \ = \ - \dot p_b
\end{equation}

The whole procedure is intended to give the following results:
The Hamiltonian $H$ shall be considered to be a usual Hamiltonian
for two interacting point particles $a(t)$ and $b(t)$. One of the
canonical equations shall be equivalent to the compatibility
condition eq. (2.6) and another one shall be equivalent to the
original Euler--Lagrange equation (2.2), whereas the two
remaining canonical equations are used to eliminate the momenta
$p_a$ and $p_b$ from the system. 
     The next step is to find those Lagrangians $L$ which make
this procedure work. 
From eqs. (2.15) and (2.11) we get
\begin{equation}
H=\frac{p_a p_b}{f(a)} - L(a,\frac{p_b}{f(a)}, b) - 
\frac{p_b \sp 2 f'(a)b}{f(a)\sp 2} - f(a)b\sp 2
\end{equation}
In this form, eq. (2.16) coincides with eq. (2.11) and (2.17)
with (2.12). So we may use eqs. (2.9, 10) in the following,
because they are equivalent to eqs. (2.11, 12). 

Now, we use eqs. (2.19), cancel  $p_b$ by use of eq. (2.10) and
get  \begin{equation}
0 = \frac{\partial L}{\partial b} + 2 b f(a) - \ddot a f(a)
\end{equation}
In order that the compatibility relation eq. (2.6) follows
automatically from eq. (2.21), one has to ensure that $f(a) \ne
0$ (which is already assumed) and that 
$$0 = \frac{\partial L}{\partial b} +  b f(a)$$
identically takes place. The condition of non--degeneracy,
$$ \frac{\partial \sp 2 L}{\partial b \sp 2} \ne 0$$
 is then also automatically fulfilled. One has the following
possible Lagrangian 
\begin{equation}
L = - \frac{1}{2} f(a) b\sp 2 + K(a, \dot a )
\end{equation}
where $K$ is an arbitrary function, but, for simplicity, we put
$K=0$.
     
The last of the four canonical equations to be used is eq. (2.18)
reading now with eqs. (2.9, 10, 20)
\begin{equation}
0 = f \ddot b + 2 f' \ \dot a \dot b 
+ \frac{3}{2} f' \ b\sp 2 + f'' \ \dot a \sp 2 b
\end{equation}
If we insert here eq. (2.5) we get exactly the same as the 
Euler--Lagrange equation (2.2) following from the Lagrangian 
\begin{equation}
L = - \frac{1}{2} f(q) (\ddot q)\sp 2
\end{equation}

Result: For every Lagran\-gian of type (2.1) which can be brought
in\-to type (2.24) with $f\ne 0$ the addi\-tion of the
diver\-gence (2.3) makes it pos\-sible to apply the new
coor\-di\-na\-tes (2.5). Then the sys\-tem be\-comes equi\-valent
to a clas\-si\-cal Ha\-mil\-tonian of two particles, and the
relation (2.6) between them follows without imposing an
additional constraint.

\subsection{A general divergence}

In this subsection we try to generalize the result of the
previous subsection by avoiding to prescribe the special
structure (2.3) of the divergence to be added. We substitute eq.
(2.3) by  
\begin{equation}
L_{div} = \frac{d}{dt} h(q, \, \dot q , \, \ddot q)
\end{equation}
Keeping eqs. (2.5) we get instead of eq. (2.7) now
\begin{equation}
\hat L = L(a, \dot a, b) + h_1 \dot a  +
h_2 b + h_3 \dot b 
\end{equation}
where $h_n$ denotes the partial derivative of $h$ with respect to
its $n$th argument. Using eqs. (2.8), (2.10) is now replaced with
\begin{equation}
p_b = h_3 (a, \dot a, b)
\end{equation}
Eq. (2.13) is kept, and (2.14) is replaced with
\begin{equation}
J = - (h_{23})\sp 2
\end{equation}
We have to require that $h_{23} \ne 0$, and then the equation
$p_b = h_3$ is locally invertible as
$\dot a  = F(p_b, a, b)$. From this definition one immediately
gets the identity $F_1 \, h_{23} = 1$. Two further identities to
be used later are not so trivial to guess. To derive them, let us
for a moment fix $p_b$ and then calculate the increase of $h_3$
with increasing $a$ and $b$ resp. The assumed constancy of $h_3$ 
yields the equations 
\begin{equation}
h_{13} \ + \ F_2 \, h_{23} \ = \ 0
\end{equation}
and
\begin{equation}
h_{33} \ + \ F_3 \, h_{23} \ = \ 0
\end{equation}
resp. 
to be used for deducing the generalization of eq. (2.21). One
gets the result: For $h_{23} \ne 0$ (which is already presumed),
the compatibility relation (2.6) follows 
automatically from the canonical  
equation (2.19) if and only if   
\begin{equation}
0 = L_3 + h_2
\end{equation}
is identically fulfilled. One can see: The condition of 
 non--degeneracy of the Lagrangian (2.1)  namely 
$$L_{33} \ne 0$$
 is equivalent to the condition $h_{23} \ne 0$. For any given 
non--degenerate Lagrangian we can find the appropriate divergence
by solving eq. (2.31) as follows
\begin{equation}
h(q, \dot q, \ddot q) = - \int _0 \sp {\dot q}
L_3(q, x, \ddot q) dx
\end{equation}
All other things are fully analogous: 
\begin{equation}
H=[p_a - h_1(a, F, b)]F - h_2(a, F, b)b - L(a, F, b)
\end{equation}
where $F=F(p_b, a, b)$. 
Eq. (2.19) with (2.30) gives the compatibility condition (2.6).
Eq. (2.18) with (2.29) is equivalent to the 
Euler--Lagrange  equation (2.2).

     \ \  Let us summarize this section: For the Lagrangian
$L=L(q, \dot q, \ddot q)$ where $\partial \sp 2 L / \partial
(\ddot q)\sp 2 \ne 0 $ we define 
 $\hat L = L + L_{div}$ where 
\begin{equation}
L_{div} \ = \  - \, \frac{d}{dt} \int \frac{\partial L}{\partial
\ddot q} (q, x, \ddot q) dx 
\end{equation}
We insert $a=q$ and $b=\ddot q$, define the momenta 
$p_a = \frac{\partial \hat L}{\partial \dot a}$ and
$p_b = \frac{\partial \hat L}{\partial \dot b}$ and get the
Hamiltonian $H = \dot a p_a + \dot b p_b - \hat L$. One of its
canonical  equations is $\ddot a = b$ and another one is
equivalent to the fourth--order Euler--Lagrange equation
following  from $L$. By these properties, $L_{div}$ is uniquely
determined up to the integration constant. Contrary to other
approaches, no constraint is needed.

\section{Instability of $R^2$-theories}

This section deals with the clas\-si\-cal insta\-bi\-li\-ty
of fourth--order theo\-ries fol\-lowing from a non--line\-ar
Lagran\-gian $L(R)$.  

Teys\-san\-dier and Tour\-renc [24], cf. also [25], 
 sol\-ved the 
Cau\-chy--prob\-lem for this theo\-ry, let us short\-ly re\-peat
the main in\-gre\-dients. 

The Cauchy problem is well--posed (a property which is usually
required to take place for a physically sensible theory) in each
interval of $R$-values where both $dL/dR$ and $d^2 L / dR^2$ are
different from zero. The constraint equations are similar as in
General Relativity: the four $0i$-component equations. What is
different are the necessary initial data to make the dynamics
unique. More exactly: Besides the data of General Relativity one
has to prescribe the values of $R$ and $\frac{dR}{dt}$ at the
initial hypersurface. This coincides whith the general
experience: Initial data have to be prescribed till the
highest--but--one temporal derivative appearing in the field
equation (here: fourth--order field equation,   $\frac{dR}{dt}$
contains third--order temporal derivatives of the metric). Under
this point of view, classical stability of the field equation
means that a small change of the Cauchy data implies also a small
change of the solution.

Now we are prepared to classify the stability claims found in
refs. [26-34]. To simplify we specialize to the Lagrangian $L=R
- \epsilon R^2$ with the non--tachyonic sign 
$\epsilon > 0$ and restrict to the range $\frac{dL}{dR} > 0$,
i.e. $R<\frac{1}{2\epsilon}$.
 On the one hand, refs. [26,27,28] 
 find a classical instability of
the Minkowski space--time for this case of fourth--order
gravity. 
Mazzitelli and Rodrigues [29] cite  ref. [30]
with the sentence ``The Minkowski solution in general relativity
has been proven to be stable.'' which refers to the positive
energy theorem of general relativity.

On the other hand, refs. [31-34] find out that the  Minkowski
space--time is not more unstable in this type of
fourth--order gravity
than in General Relativity itself. What looks like a
contradiction from the first glance is only a notational
ambivalence as can be seen now:
 The main argument in refs. [26-29]
 is that an arbitrarily  large value $\frac{dR}{dt}$ is
compatible with small values of $H^2$ and $R^2$. In refs. 
[31-34] however, following the Cauchy--data argument [24,25],
($\frac{dR}{dt}$ being part of the Cauchy data which are presumed
to be small) stability of the Minkowski space--time is obtained
in the version: If the Cauchy data are small (meaning: close to
the Cauchy data of the Minkowski space--time) then the
fourth--order field equation bounds the solution to remain close
to the Minkowski space--time.

The argument of ref. [31] is a little bit different: There the
conformal transformation to Einstein's theory with a scalar field
$\Phi$ [20] is applied; it is observed that in the $F(R)$-theory
there are never ghosts which implies stability. 
Now, $\Phi$ and $\frac{d\Phi}{dt}$ belong to the Cauchy data
which is equivalent to the data $R$, $\frac{dR}{dt}$ in the
conformal picture thus supporting the Cauchy data argument given
at the beginning of this section.

\section{Cosmology}
\setcounter{equation}{0}

Several papers [35-43] apply the conformal transformation
theorem [20] to cosmology; so for interpreting the  
cosmological singularity [35], for dealing with
 anisotropic models [36,37], with transformation to Brans--Dicke
extended inflation [38]. 
Other papers  apply this theorem  as a mathematical device
to transform exact solutions 
of one of the theories 
to solutions of the other theory. 

 Chimento [44] found an exact solution for fourth--order
gravity in a spatially flat Friedmann model. He also found out
that in the tachyonic--free case the asymptotic matter-dominated
Friedmann solution is stable, and no fine--tuning of initial
conditions is necessary to get the final (oscillating) Friedmann
stage; particle production of non--conformal fields may backreact
to damp the oscillations. 
[45] generalizes [44]: here the Dirac equation is considered, the
result is that there appear also spinor field oscillations.

Let us present the exact solution of [44]. For the spatially flat
Friedmann model with Hubble parameter 
$H=\dot a/a$ he solves the fourth--order field equation with
vacuum polarization term.
The zero--zero component equation reads
\begin{equation}
2H\ddot H - \dot H^2 + 6 H^2\dot H + \frac{9}{4} H^4 + H^2=0
\end{equation}
The $H^4$-term stems from the vacuum polarization and the
$H^2$-term from the Einstein tensor. The remaining ingredients of
eq. (4.1) come from the term $R^2$ in the Lagrangian. (Here we
only present the tachyonic--free case with $\Lambda =0$ and
$\frac{9}{4}$ in front of $H^4$.) The factor in front of $H^4$
should not influence the weak--field behaviour because for
$H\approx 0$ this factor only changes the effective gravitational
constant.

From eq. (4.1) the discussion of section 3 becomes obvious:
(4.1) represents a third--order equation for the cosmic scale
factor $a$; it is a constraint and not a dynamical equation. (It
is only due to the high symmetry, that accidentally the validity
of the constraint implies the validity of the dynamical
equation.) Supposed, eq. (4.1) would be  the true dynamical
equation for a theory, then the instability argument of 
[26-28] could apply. 

 The ansatz for solving eq. (4.1) 
$$H=\frac{2\dot s}{3s}$$
leads to a non--linear third--order equation for $s$
\begin{equation}
2 \dot s s^{(3)} - \ddot s ^2 + \dot s^2 =0
\end{equation}
Derivative with respect to $t$ yields  
the  equation $s^{(4)} + \ddot s=0$
being linear in $s$ and having the solution 
$$s= c_1 + c_2 t + c_3 \sin (t + c_4)$$
Inserting this solution into the original equation gives the
restriction $\vert c_2\vert = \vert c_3\vert$. 
Let us discuss this solution: 
$c_2=0$ leads to the uninteresting flat space--time. 
So, now let $c_2 \ne 0$. Adding $\pi$ to $c_4$ can be absorbed by
a change of the sign of $c_3$. Therefore, $c_2=c_3$ without loss
of generality. Multiplication of $s$ by a constant factor does
not change the geometry, so let $c_2=1$. A suitable
time--translation leads to $c_1=0$. Finally, the cosmic scale
factor is calculated as $a=s^{2/3}$ leading to
\begin{equation}
a\ = \ [ t +  \sin (t + c_4) ]
^{2/3} \ \sim \ t^{2/3} \, [1 + \frac{2}{3t} \sin (t + c_4) ]
\end{equation}
The r.h.s. of eq. (4.3) gives  the late--time
behaviour  deduced in [46]. The factor $1/t$ in front of
the  ``sin''-term shows that the oscillations due to the
higher--order terms are damped. The total energy ``sitting''
 in these oscillations, however, remains constant in time
(because of the volume--expansion), cf.  [27], and can be
converted into classical matter by particle creation.

Let us mention some further cosmological solutions with
higher--order gravity: [47] discusses the $L(R)$-stability with a
conformally coupled scalar field. Ref.
[48] (partial results of it can be found in [49]) deals with 
fourth--order cosmological models of Bianchi--type I and 
power--law metrics, i.e.
$$ds^2 = dt^2 - \sum_{i=1} ^3 \ t^{2p_i} \ (dx^i)^2 $$
with real parameters $p_i$. The suitable notation
$$a_k =  \sum_{i=1} ^3 \ p_i ^k$$ 
gives the following: $a_1=a_2=1$ is the usual Kasner solution for
Einstein's theory. $a_1^2 + a_2 = 2 a_1$ is the condition to be
fulfilled for a solution in $L=R^2$. Refs. [50, 51] also  
discuss  $R^2$-models. Ref. [33]
considers inflationary cosmology with a Lagrangian  
$$ L \ = \  R  + \ \lambda R_{\mu \nu} R^{\mu \nu}/R.$$ 
Ref. [52] deals with anisotropic Bianchi--type IX solutions for
$L=R^2$. They look for chaotic behaviour analogous to the
mixmaster model in Einstein's theory. Ref. [53] gives exact
solutions for $L=R^2$ and a closed Friedmann model, ref. 
[54] discusses the bounce in closed Friedmann models for $L=R - 
\epsilon R^2$. Supplementing the discussion of [54, eq.(1)] let
us mention: In the non--tachyonic case, there exist periodically
oscillating models with an always positive scale factor $a$. Ref.
[55] looks for chaos in isotropic models, e.g. by conformally
coupled massive scalar fields in the closed universe. The papers
[56,57] consider the stability of power--law inflation for
$L=R^m$ within the set of spatially flat Friedmann models. Refs.
[58] give overviews on higher--order cosmology, especially
chaotic inflation as an attractor solution in initial--condition
space. [59] deals with quantum gravitational effects in the de
Sitter space--time, and [60] gives a classification of 
in\-fla\-tio\-na\-ry Ein\-stein--sca\-lar--field--mo\-dels via
ca\-ta\-stro\-phe theory. Ref. [61] considers  Chern--Simon terms
in Bianchi cosmologies and the cosmic no-hair conjecture.

\section{Summary}
\setcounter{equation}{0}

The scope of this paper was to review recent work 
 connected with higher--order theories, especially
 fourth--order gravity theories and their 
application to cosmology. We 
 presented a step necessary to deduce the 
Wheeler--de Witt equation for a cosmological 
minisuperspace model in fourth--order gravity
 and discussed the several stability claims and 
conformal transformation theorems.

The method (first used in [51] for $L=R^2$ and a
spatially flat Friedmann model) to handle with eqs. (1.3 - 1.6)
was systematically generalized in sct. 2 (cf. also [62]) 
to give a Hamiltonian
formulation of a general fourth--order theory. The possibility of
deducing this method makes it clear that the method of ref. [51]
is not restricted to highly symmetric models. The alternative
Hamiltonian formulation has some advantages in comparison with
Ostrogradski's one: No constraint is needed, the Hamiltonian is
typically a quadratic function in the momenta. (Ostrogradski's
approach leads always to a Hamiltonian linear in the momenta
which gives artificial factors $i$ in the Schr\"odinger
equation.) The calculation of the momenta  from the Lagrangian
follows the usual equations (2.8)  whereas the Ostrogradski
approach needs some additional terms. Our approach is less
ambiguous, cf. eq. (2.7).

The fact that the Jacobian eq. (2.28) is always negative excludes
the possibility to get a positive definite Jacobi metric in eq.
(1.2). This is one of the many possibilities to say what is meant
by the phrase ``fourth--order theories are always unstable''.
 The
Jacobi metric plays the role of the conformally transformed
superspace--metric used in quantum cosmology. And here the circle
can be closed: In Einstein's theory (both for Lorentzian and 
Euclidean signature of the underlying manifold) the
superspace--metric has Lorentzian signature and cannot be
positive definite. So we get once more the result: 
Fourth--order gravity contains 
 instabilities,  but for the non--tachyonic 
 case of $L=R-\epsilon R^2$ 
only those which it has in common with General Relativity. 

\medskip

Let us finish by mentioning some open
questions worth being attacked in the future: 

\medskip

For other types of fourth-- and higher--order
theories the singularity behaviour is not yet understood.

To decide the quantum instability of the Minkowski or de Sitter
space--times in fourth--order gravity one must solve the
corresponding Wheeler--de Witt equations. 
In [29] they are deduced for the spatially
flat Friedmann model and the Lagrangian $L=R-\epsilon R^2$. 

\medskip

In [50] it is mentioned that a classical theory with higher
derivatives has instabilities: ``At the quantum level, the
difference is even more dramatic. Noncommuting variables in the
lower--derivative theory, such as position and velocities, become
commuting in the higher--derivative theory.''
 Remark of U. Kasper
to this sentence: ``The uncertainty relation is primarily between
positions and momenta. If the momentum is independent of the
velocity then commuting position and velocity need not bother.''

\medskip

Refs.  [62-64] 
discuss the conformal transformations between fourth--order
gravity to Einstein's theory with a scalar field.
Reuter [62] proposed to use the notion ``Bicknell--theorem''
to  state that conformal equivalence. Dick [64] 
represents  a valuable update of the discussion which 
 of the two conformally equivalent metrics shall be 
considered physical.

\medskip

Because of the frequent appearance of 
singular points of the corresponding differential 
equations it is especially
useful to have a growing set of 
exact solutions (see also [65])
 of the non-linear gravity models.

\bigskip

{\it Acknowledgement}. I thank V. Ivashchuk, 
 U. Kasper, V. Melnikov and  M. Rainer
for elucidating comments. 
 Financial support from the  Deutsche
For\-schungs\-gemein\-schaft  is gratefully
acknowledged.

\bigskip

{\Large {\bf References}}

\noindent [1] H.-J. Schmidt, Gen. Relat. Grav. {\bf 29}, 859
(1997).

 \medskip 
\noindent [2] M. Ostrogradski, Memoires Academie St. Petersbourg
Series
{\bf VI} vol.  {\bf 4},  385 (1850); D. Eliezer 
and R. Woodard, Nucl. Phys. {\bf B 325}, 389 (1989).

 \medskip 
\noindent [3] X. Jaen, J. Llosa and A. Molina, Phys. Rev. {\bf D
34},
2302 (1986); J. Llosa and J. Vives, Int. J. Mod. Phys. 
{\bf D 3}, 211 (1994).

 \medskip 
\noindent [4] J. Govaerts, Hamiltonian quantization and
constrained
dynamics, Leuven Univ. Press 1991; 
 J. Govaerts, Phys. Lett. {\bf B 293}, 327 (1992);
 U. Kasper, Class. Quantum Grav. {\bf 10}, 869 (1993); 
 U. Kasper, Gen. Relat. Grav. {\bf 29}, 221 (1997).

 \medskip 
\noindent [5] G. Kleppe, Ana\-ly\-sis of higher deri\-va\-tive
hamil\-tonian for\-ma\-lism, Pre\-print UAHEP-9407 Univ. of
Alabama 1994, representing a comment to [51].

 \medskip 
\noindent [6] C. Battle, J. Gomis, J. Pons and N. Roman--Roy, J.
Phys.
{\bf A 21}, 2693 (1988); V. Nesterenko, J. Phys. {\bf A 22}, 1673
(1989).

 \medskip 
\noindent [7] X. Gracia, J. Pons and N. Roman--Roy, J. Math.
Phys. 
{\bf 32}, 2744 (1991).

 \medskip 
\noindent [8] A. Chervyakov and V. Nesterenko, Phys. Rev. {\bf D
48},
5811 (1993).

 \medskip 
\noindent [9] T. Damour and G. Sch\"afer, J. Math. Phys. {\bf
32}, 127
(1991); V. Perlick,  J. Math. Phys. {\bf 33}, 599 (1992).

 \medskip 
\noindent [10] J. Douglas, Trans. Amer. Math. Soc. (TAMS) {\bf
50}, 71
(1941).

 \medskip 
\noindent [11] M. Szydlowski and J. Szczesny, Phys. Rev. {\bf D
50}, 819
(1994).

 \medskip 
\noindent [12] K. Stelle, Gen. Relat. Grav. {\bf 9}, 353 (1978).

 \medskip 
\noindent [13] L. Bel and H. Sirousse-Zia, Phys. Rev. {\bf D 32},
3128
(1985).

 \medskip 
\noindent [14] J. Navarro-Salas, M. Navarro, C. Talavera and V.
Aldaya,
Phys. Rev. {\bf D 50}, 901 (1994).

 \medskip 
\noindent [15] P. Teyssandier, Class. Quant. Grav. {\bf 6}, 219 
(1989).

 \medskip 
\noindent 
[16]  P. Teyssandier, Annales de Physique,
  Coll.1, Suppl.6, {\bf 14}, 163  (1989).

 \medskip 
\noindent 
[17] P. Teyssandier, Astron. Nachr. {\bf 311}, 209  (1990).

 \medskip 
\noindent [18] B. Linet and P. Teyssandier, Class. Quant. Grav.
{\bf 9},
159  (1992).

 \medskip 
\noindent [19] R. Wagoner, Phys. Rev. {\bf D 1},
3209 (1970).

 \medskip 
\noindent [20] M. Baibosunov, V. Gurovich and U. Imanaliev,
 J. eksp. i teor. fiz. {\bf 98}, 1138  (1990)
 and ref. cited there.

 \medskip 
\noindent [21] J. Audretsch, A. Economou and C. Lousto, Phys.
Rev.    
 {\bf D 47}, 3303 (1993).

 \medskip 
\noindent 
[22] H.-J. Schmidt, Phys. Rev.    
 {\bf D 52} (1995) 6198.

 \medskip 
\noindent [23] G. Magnano and L. Soko\l owski, Phys. Rev. {\bf D
50}, 5039
(1994).

 \medskip 
\noindent [24] P. Teyssandier and Ph. Tourrenc, J. Math. Phys.
{\bf 24},
2793 (1983).

 \medskip 
\noindent 
[25] A.A. Starobinsky and H.-J. Schmidt, Class. Quant.
Grav. {\bf 4}, 695  (1987).

 \medskip 
\noindent [26] W.-M. Suen, Phys. Rev. {\bf D 40}, 315 (1989);

 \medskip 
\noindent [27] 
W.-M.  Suen,
Phys. Rev. {\bf D 50}, 5453 (1994);

 \medskip 
\noindent 
[28] D. Coule and M. Madsen, Phys.
Lett. {\bf B 226}, 31 (1989).

 \medskip 
\noindent [29] F. Mazzitelli and L. Rodrigues, Phys. Lett. {\bf B
251},
45 (1990).

 \medskip 
\noindent [30] D. Gross, M. Perry, L. Yaffe, Phys. Rev. {\bf D
25}, 330
(1982).

 \medskip 
\noindent [31] G. Lopez Cardoso and B. Ovrut, Int. J. Mod. Phys.
{\bf D
3}, 215 (1994).

 \medskip 
\noindent [32]  V. M\"uller, Int. J. Mod. Phys. {\bf D 3}, 241
(1994).

 \medskip 
\noindent 
[33] 
 E. Br\"uning, D. Coule and C. Xu, Gen. Relat. Grav. {\bf 26},
1197 (1994).

 \medskip 
\noindent 
[34] H.-J. Schmidt, Phys. Rev. {\bf D 50}, 5452 (1994).

 \medskip 
\noindent [35] G. Le Denmat and H. Sirousse-Zia, Phys. Rev. {\bf
D 35},
480 (1987).

 \medskip 
\noindent [36] L. Pimentel and J. Stein - Schabes, Phys. Lett.
{\bf B 
216}, 27 (1989).

 \medskip 
\noindent [37] O. Bertolami, Phys. Lett. {\bf B 234}, 258 
(1990).

 \medskip 
\noindent [38] Y. Wang, Phys. Rev. {\bf D 42}, 2541  (1990).

 \medskip 
\noindent [39] J. Barrow and K. Maeda, Nucl. Phys. {\bf B 341},
294 (1990).

 \medskip 
\noindent 
[40] D. Hochberg, Phys. Lett.  {\bf B 251}, 349  (1990).

 \medskip 
\noindent [41] D. La, Phys. Rev. {\bf D 44}, 1680 (1991).

 \medskip 
\noindent [42] U. Bleyer, Proceedings Conference Physical
Interpretations
of Relativity Theory, London 1992, p. 32;

 \medskip 
\noindent 
[43]  S. Mignemi and D. Wiltshire, Phys. Rev. {\bf D 46}, 1475  
(1992);  L. Amendola, D. Bellisai, R. Carullo and F. Occhionero,
p. 127 in:  Relativistic  Astrophysics and Cosmology,  Eds.:  S.
Gottl\"ober, J. M\"ucket, V. M\"uller WSPC Singapore 1992;  S.
Capozziello, L. Amendola and F. Occhionero, ditto  p. 122.

 \medskip 
\noindent [44] L. Chimento, Class. Quant. Grav. {\bf 5}, 1137
(1988); 
 L. Chimento, Class. Quant. Grav. {\bf 6}, 1285 (1989);
 L. Chimento, Class. Quant. Grav. {\bf  7}, 813 (1990); L.
Chimento, Gen. Relat. Grav. {\bf 25}, 979  (1993).

 \medskip 
\noindent [45] L. Chimento, A. Jakubi and F. Pensa, Class. Quant.
Grav.
{\bf 7}, 1561 (1990).

 \medskip 
\noindent [46] V. M\"uller and H.-J. Schmidt, Gen. Relat. Grav.
{\bf 17},
769  (1985).

 \medskip 
\noindent [47] C. Laciana, Gen. Relat. Grav. {\bf 25}, 245     
(1993).

 \medskip 
\noindent [48] H. Caprasse, J. Demaret, K. Gatermann and H.
Melenk,
 Int. J. Mod. Phys. {\bf C 2}, 601  (1991).

 \medskip 
\noindent [49] V. M\"uller, Ann. Phys. (Leipz.) {\bf 43}, 67
(1986).

 \medskip 
\noindent [50] F. Mazzitelli, Phys. Rev. {\bf D 45}, 2814 (1992).

 \medskip 
\noindent [51] H.-J. Schmidt, Phys. Rev. {\bf D 49}, 6354 (1994),
Erratum {\bf D 54}, 7906 (1996).

 \medskip 
\noindent [52] P. Spindel and M. Zinque, Int. J. Mod. Phys. {\bf
D 2}, 
279  (1993).

 \medskip 
\noindent [53] H.-J. Schmidt, Gen. 
Relat. Grav. {\bf 25}, 87  (1993), Erratum p. 863.

 \medskip 
\noindent [54] D. Coule, Class. Quant. Grav. {\bf 10}, L25 
(1993).

 \medskip 
\noindent [55] E. Calzetta, Int. J. Mod. Phys. {\bf D 3}, 167
(1994).

 \medskip 
\noindent [56] A. Burd and J. Barrow, Nucl. Phys. {\bf B 308},
929  (1988);
C. Sivaram and V. de Sabbata, p. 503 in: Quantum Mechanics in
Curved  Space-Time, Eds. J. Audretsch, V. de Sabbata, Plenum NY
1990; V.  M\"uller, H.-J. Schmidt and A.A. Starobinsky, Class.
Quant. Grav. {\bf 7}, 1163  (1990);
 Y. Kitada and K. Maeda, Phys. Rev. {\bf D 45}, 1416  (1992);
 J. Aguirregabiria, A. Feinstein and J. Ibanez, Phys. Rev. {\bf D
48}, 4662 (1993);  J. Aguirregabiria, A. Feinstein and J. Ibanez,
Phys. Rev. {\bf D 48}, 4669 (1993).

 \medskip 
\noindent [57] S. Cotsakis and P. Saich, Class. Quant. Grav.
 {\bf 11}, 383 (1994).

 \medskip 
\noindent [58] J. Barrow, Lect. Notes Phys. {\bf 383}, 1  (1991);
 J. Kung and R. Brandenberger, Phys. Rev. {\bf D 42},
1008 (1990); S. Kluske, Dissertation Potsdam University 1997.

 \medskip 
\noindent [59] C. Kiefer, 203  in: 
 New Frontiers in Gravitation, ed. G. Sardanashvily 
 (Hadronic Press, 1996).

 \medskip 
\noindent [60] F. Kusmartsev, E. Mielke, Y. Obukhov and
F. Schunck, Phys. Rev. {\bf D 51}, 924 (1995).

 \medskip 
\noindent [61] N. Kaloper, Phys. Rev. {\bf D 44}, 2380 (1991).

\medskip
\noindent [62] S. Reuter, Dissertation Potsdam University 1997.

\medskip
\noindent [63] H. v. Elst, Dissertation QMW College London 1996, 
esp. Chapter 6.

\medskip
 \noindent [64] R. Dick, Gen. Relat. Grav. {\bf 30} (1998)
 in print, and references cited there.

\medskip
 \noindent [65] S. Chervon, V. Zhuravlev, V. Shchigolev,
Phys. Lett. {\bf B 398}, 269 (1997).

\end{document}